\documentclass[preprint, superscriptaddress, 
amsmath,amssymb, aps, prb, citeautoscript]{revtex4-2}

\usepackage{graphicx}
\usepackage{dcolumn}
\usepackage{bm}
\usepackage{xcolor}
\usepackage[normalem]{ulem}

\begin{document}

\title{Resolving exciton and polariton multi-particle correlations in an optical microcavity in the strong coupling regime}

\author{Victoria~Quir\'os-Cordero}
\altaffiliation{V.Q.C.\ and E.R.G.\ are first co-authors}
\affiliation{School of Materials Science and Engineering, Georgia Institute of Technology, Atlanta, GA~30332, United~States}

\author{Esteban~Rojas-Gatjens}%
\altaffiliation{V.Q.C.\ and E.R.G.\ are first co-authors}
\affiliation{School of Chemistry and Biochemistry, Georgia Institute of Technology, Atlanta, GA~30332, United~States}%
\affiliation{School of Physics, Georgia Institute of Technology, Atlanta, GA~30332, United~States}%

\author{Martin~Gomez-Dominguez}%
\affiliation{School of Materials Science and Engineering, Georgia Institute of Technology, Atlanta, GA~30332, United~States}

\author{Hao~Li}
\affiliation{Institut Courtois \& D\'epartement de Physique, Universit\'e de Montr\'eal, Montr\'eal H2V~0B3, Qu\'ebec, Canada}

\author{Carlo~A.~R.~Perini}
\affiliation{School of Materials Science and Engineering, Georgia Institute of Technology, Atlanta, GA~30332, United~States}

\author{Natalie~Stingelin}
\affiliation{School of Materials Science and Engineering, Georgia Institute of Technology, Atlanta, GA~30332, United~States}
\affiliation{School of Chemical and Biochemical Engineering, Georgia Institute of Technology, Atlanta, GA~30332, United~States}

\author{Juan-Pablo~Correa-Baena}
\affiliation{School of Materials Science and Engineering, Georgia Institute of Technology, Atlanta, GA~30332, United~States}

\author{Eric~R.~Bittner}
\affiliation{Department of Physics, University of Houston, Houston, TX~77204, United~States}

\author{Ajay~Ram~Srimath~Kandada}
\email{srimatar@wfu.edu}
\affiliation{Department of Physics \& Center for Functional Materials, Wake Forest University, Winston-Salem, NC~27109, United~States}

\author{Carlos~Silva-Acu\~na}
\email{carlos.silva@umontreal.ca}
\affiliation{School of Chemistry and Biochemistry, Georgia Institute of Technology, Atlanta, GA~30332, United~States}
\affiliation{Institut Courtois \& D\'epartement de Physique, Universit\'e de Montr\'eal, Montr\'eal H2V~0B3, Qu\'ebec, Canada}

\date{\today}

\begin{abstract}
Multi-particle correlations of exciton-polaritons and reservoir-excitons in the strong light-matter coupling regime dictate the quantum dynamics of optical microcavities. In this work, we examine the many-body exciton-polariton dynamics in a Fabry-P\'erot microcavity of a two-dimensional metal-halide semiconductor over timescales involving polariton ($\ll 1$\,ps) and exciton ($\gg 1$\,ps) scattering. We find enhanced exciton nonlinear dynamics in the microcavity versus the bare semiconductor, concomitant with ultrafast polariton scattering dynamics. We measure, by means of coherent spectroscopy, the coupling between exciton-polaritons, bright excitons, and reservoir-excitons that highlight the complex scattering landscape that fundamentally drives polariton condensation.
\end{abstract}

\maketitle

\section{Introduction}

\begin{figure}[b!]
    \centering
    \includegraphics[width=8.6cm]{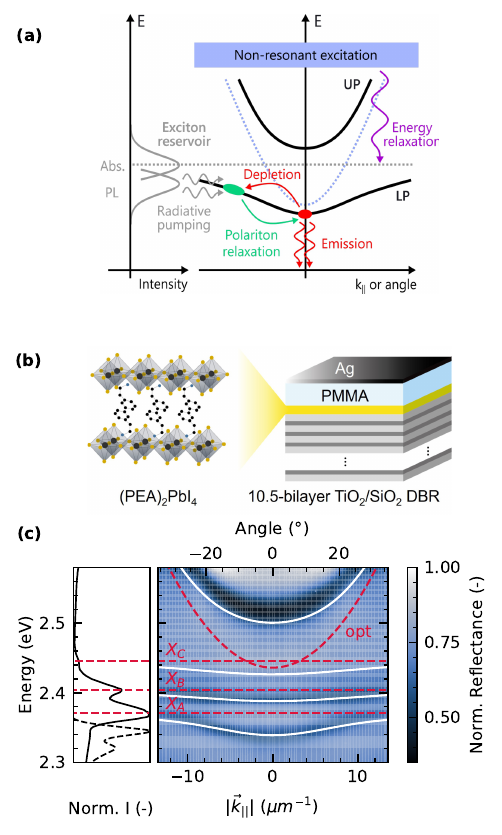} 
    \caption{(a) Schematic representation of competing mechanisms populating and depopulating the lower-polariton $\vec{k}_\parallel=\vec{0}$ state (LP: lower polariton, UP: upper polariton).
    (b) Fabry-P\'erot microcavity consisting of a distributed Bragg reflector (10.5 quarter-wave bilayers of TiO$_2$/SiO$_2$, central wavelength 520\,nm), a 60-nm (PEA)$_2$PbI$_4$ film, a 125-nm poly(methylmethacrylate) spacer layer, and a 40-nm Ag film.
    (c) The reflectance energy dispersion at 5\,K (right panel). The left panel includes the absorption (solid line) and photoluminescence (dashed line) spectra of (PEA)$_2$PbI$_4$.
    }
    \label{fig:cartoon_and_kspace}
\end{figure}

Exciton polaritons are hybrid light-matter quasiparticles arising from the strong coupling between excitons and cavity photons, notable for their ability to undergo Bose-Einstein-like condensation~\cite{Kasprzak2006, BaliliSnoke2007_CondensateTrap, polimeno2020observation}. Such condensates provide a unique platform to explore non-equilibrium many-body physics, offering insight into spontaneous coherence in driven-dissipative systems that lie beyond the scope of traditional equilibrium condensates. The collective dynamics of exciton polaritons are predominantly governed by microscopic energy redistribution mechanisms that funnel the population into the lowest in-plane momentum state ($\Vec{k}_\parallel=\Vec{0}$), often through a complex interplay of momentum-conserving scattering pathways; see Fig.~\ref{fig:cartoon_and_kspace}(a). These may include polariton–polariton parametric scattering and interactions with a reservoir of optically dark excitons. At their core, these processes are driven by many-body interactions mediated by multi-exciton correlations~\cite{Schwendimann2008, Pieczarka2020} and the strength of the cavity-enhanced inter-excitonic Coulomb potential~\cite{Takemura2016, Takemura2016Dephasing, fieramosca2024electron}.  

Exciton-polariton dynamics are typically measured by time-resolving emission and transient absorption spectra, but these measurements tend to reflect the population in the exciton reservoir. By the reservoir, we do not mean bare excitons, but rather the dark states that arise from the coupling of many exciton
states to the optical mode. These normally dark states are coupled to the bright states due to the inhomogeneous bandwidth. 
In this context, coherent nonlinear spectroscopy probes directly phase-coherent nonlinear response functions that encode the temporal evolution of excited-state coherences and interaction-induced couplings beyond population dynamics within the polariton lifetime. In this work, we employ a two-dimensional metal halide (2DMH) semiconductor platform to explore polariton many-body dynamics. Prototypical 2DMHs such as (PEA)$_2$PbI$_4$  (PEA = phenylethylammonium) feature large exciton binding energies ($> 200$\,meV), high oscillator strengths, and minimal Stokes shifts. Crucially, these materials also exhibit pronounced multi-exciton correlations~\cite{Thouin2018Bi, FelixEnhancedScreening, Srimath2020Stochastic, gatjens2023manyexciton}, positioning them as ideal candidates for studying interaction-driven exciton-polariton dynamics beyond the mean-field regime. Significantly, in bare (PEA)$_2$PbI$_4$ films of comparable thickness as the active layer in Fig.~\ref{fig:cartoon_and_kspace}(b), the homogeneous and inhomogeneous spectral linewidths are comparable ($\sim2$--8\,meV, depending on the exciton density, versus $\sim 6$\,meV, respectively)~\cite{FelixEnhancedScreening}, in contrast to organic semiconductors, which also exhibit strongly nonlinear polariton dynamics~\cite{RadPumpCarlosLidzey, Deshmukh:23, Laitz2023, Margkou2010, Somaschi2011, Coles2011}, but in which inhomogenous broadening effects might dominate in microcavities~\cite{Houdre1996,DarknessMusser}.

\section{Nonlinear exciton-reservoir dynamics}

A schematic of the (PEA)$_2$PbI$_4$ microcavity is included in Fig.~\ref{fig:cartoon_and_kspace}(b). 
The reflectance energy dispersion shows four bands corresponding to an upper, two middle, and a lower polariton branch (Fig.~\ref{fig:cartoon_and_kspace}(c), right panel). 
The multiple polaritons arise from the coupling of the microcavity optical mode (opt) with three excitonic features of (PEA)$_2$PbI$_4$~\cite{Neutzner20181ExcitonPolaron, thouin2019phonon, Srimath2020polaron, Straus2022, Dysik2024PolaronVibronic}, clear in the low-temperature absorption spectrum ($X_\text{A}$, $X_\text{B}$, and $X_\text{C}$ in Fig.~\ref{fig:cartoon_and_kspace}(c), left panel). The dispersion agrees with the eigenstates of a Hamiltonian in which the three excitons couple independently to a unique optical mode (white solid lines in Fig.~\ref{fig:cartoon_and_kspace}(c))~\cite{SM}. The theoretical microcavity quality factor is $Q \approx 68$, which corresponds to an optical mode lifetime $\tau_\text{opt} \approx 37$\,fs and a LP lifetime $\tau_\text{LP} \approx 114$\,fs~\cite{SM}. This is consistent with the width of the LP branch in Fig.~\ref{fig:cartoon_and_kspace}(c) from which we estimate a polariton lifetime of 165\,fs. This implies that measurements on picosecond timescales resolve the exciton reservoir dynamics imprinted in the polariton dispersion, and not the polariton dynamics, which evolve on $\lesssim 100$\,fs.

\begin{figure}[b!]
    \centering
   \includegraphics[width=8.6cm]{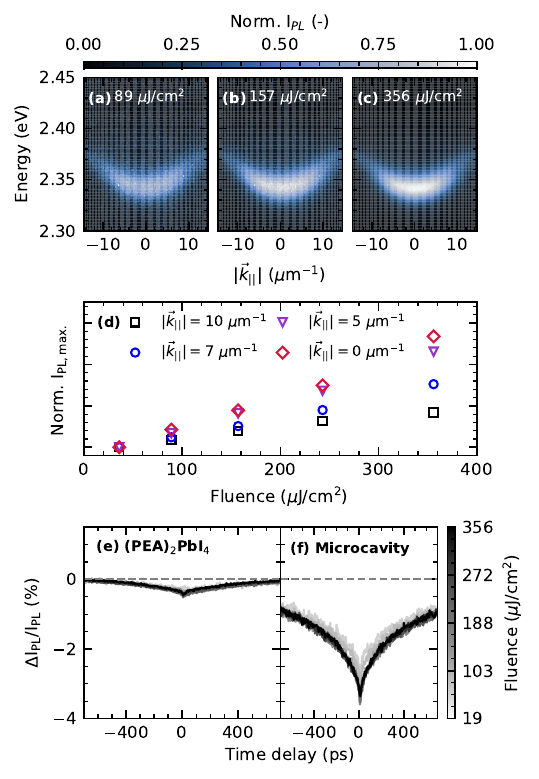} 
    \caption{(a-c) Normalized photoluminescence energy dispersion at distinct incident excitation fluences. 
    (d) Fluence dependence of the maximum PL intensity ($I_\text{PL,max}$) for various $\vec{k}_{\parallel}$. (e-f) Fraction of nonlinear PL ($\Delta$I$_\mathrm{PL}$/I$_\mathrm{PL}$), measured via ECPL, in a bare (PEA)$_2$PbI$_4$ film and the microcavity. The PL energy dispersion and the ECPL experiments share the same non-resonant excitation (2.638\,eV, 200\,fs) and collection conditions; however, for the ECPL measurements, we integrate the PL of the entire lower polariton branch.}    
    \label{fig:Nonlinearities}
\end{figure}

Under non-resonant excitation, we observe that the photoluminescence (PL) arises predominantly from the LP with  $|\Vec{k}_\parallel|<\,10\,\mu$m$^{-1}$. Fig.~\ref{fig:Nonlinearities}(a--c).
High wavevectors show a clear sublinearity in the maximum PL intensity ($I_\text{PL,max}$, Fig.~\ref{fig:Nonlinearities}(d)).

We identify the exciton reservoir nonlinear dynamics, imprinted in the emission of the lower polariton ($\gg 1$\,ps), and compare them to those of excitons in a bare (PEA)$_2$PbI$_4$ film.
For this purpose, we use excitation correlation PL (ECPL) spectroscopy~\cite{Johnson_1988, Srimath2016_ECPL, Rojas2023ECPL}. In this experiment, the fractional change in PL intensity $\Delta I_\mathrm{PL}$ of the total intensity $I_\mathrm{PL}$ due to linear excitation by two variably delayed pulses, arising from nonlinear interaction of two photoexcitations, quantifies the dimensionless signal $\Delta I_\mathrm{PL}/I_\mathrm{PL}$. This quantity can be meaningfully compared for photoexcitations in the bare semiconductor and within the microcavity, as the amplitude measures the contribution of nonlinear population dynamics to the total PL intensity~\cite{Rojas2023ECPL}. 
In the bare (PEA)$_2$PbI$_4$ film, we measure a small negative signal for all time delays and incident fluences, $|\Delta I_\mathrm{PL}/I_\mathrm{PL}| < 0.5\%$ in Fig.~\ref{fig:Nonlinearities}(e). 
In the microcavity, we observe a negative nonlinear contribution that is one order of magnitude greater than that measured in the film, $|\Delta I_\mathrm{PL}/I_\mathrm{PL}| < 4\%$ in Fig.~\ref{fig:Nonlinearities}(f). 
The magnitude of the negative nonlinear PL signal is indicative of the extent of exciton-exciton annihilation in each system~\cite{Rojas2023ECPL, SM}, suggesting increased interactions in the exciton reservoir compared to excitons in (PEA)$_2$PbI$_4$.
We rationalize the exciton-exciton annihilation increase in the microcavity due to the spatial delocalization of the exciton reservoir via population exchange with polaritons. 
This hypothesis goes in line with recent work reporting a rapid exchange between polaritons and exciton-polariton transport at timescales beyond the polariton lifetime driven by the exciton reservoir~\cite{Groenhof2019, RaoMusserReservoirDeloc, Xu2023, Jin2023_Enhanced}.

\section{Polariton spectral correlations}

\begin{figure}[b!]
 \centering
    \includegraphics[width=8.6cm]{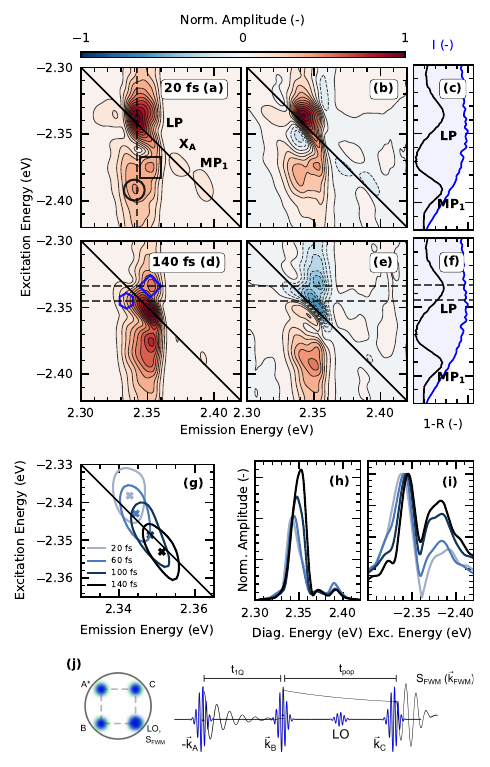}
    \caption{1Q rephasing spectra at 10\,K, where the absolute and real components at population times ($t_{\mathrm{pop}}$) of 20\,fs and 140\,fs are shown in (a-b) and (d-e), respectively. The dashed line corresponds to a cut along the emission energy axis at 2.342\,eV. 
    (c-f) Laser spectrum superposed with the normalized (1-R) of the microcavity at $\Vec{k}=\Vec{0}$.
    The population evolution of the absolute component is summarized by tracking the (g) contour of the LP, (h) the cut along the diagonal, and (i) the vertical cut marked in (a).
    (j) In the experimental setup, the pulses are arranged in a \textit{BoxCARS} geometry (left) and time-ordered for phase-matching corresponding to the 1Q rephasing spectra (right).
    The pulses excite the sample at an angle of 2.3\,$^\circ$, which corresponds to $|\vec{k}_\parallel | =0.92\,\mu\mathrm{m}^{-1}$.}
    \label{fig:1Q}
\end{figure}

We explore the polariton nonlinear dynamics by using coherent two-dimensional spectroscopy. We show, in Fig.~\ref{fig:1Q}, the one-quantum (1Q) rephasing spectra~\cite{SM}. 
The $\sim 20$\,fs pulses are resonant with LP and MP$_\mathrm{1}$ branches at small incidence angle, as well as with the $X_\text{A}$ exciton reservoir (Fig.~\ref{fig:1Q}(c)). At early population times ($t_\mathrm{pop} = 20$\,fs), we observe three main features along the diagonal with energies of 2.342, 2.371, and 2.390\,eV (Figs.~\ref{fig:1Q}(a-b)) assigned to LP, $X_\text{A}$, and MP$_\mathrm{1}$, respectively, in correspondence with the reflectance of the microcavity at $\vec{k}_\parallel \sim \Vec{0}$, Fig.~\ref{fig:1Q}(c), and the absorption of (PEA)$_2$PbI$_4$ (Fig.~\ref{fig:cartoon_and_kspace}(c)). 
The clear feature corresponding to $X_\text{A}$ can be rationalized through exciton-polariton models that include disorder (i.e.\ inhomogeneous broadening, $\sim 6$\,meV in the bare film~\cite{FelixEnhancedScreening}) in the exciton reservoir~\cite{Houdre1996, DarknessMusser} and has also been seen in two-dimensional coherent spectroscopy measurements of a strongly coupled organic microcavity~\cite{Mewes2020}.

The real component of the 1Q rephasing spectra uncovers the phase behavior of the LP, $X_\text{A}$, and MP$_\mathrm{1}$  signatures. 
At $t_\mathrm{pop} = 20$\,fs (Fig.~\ref{fig:1Q}(b)), $X_\text{A}$ and MP$_\mathrm{1}$ show dispersive lineshapes, with a $\pi$-shift difference between them.
The lineshape of the LP looks more complex, possibly due to overlapping transitions \cite{gatjens2023manyexciton, Son2022}.
As population time advances to $140$\,fs, Fig.~\ref{fig:1Q}(e), the lineshape of the LP signature declutters, and we observe a clear dispersive lineshape with the same phase as that of MP$_\mathrm{1}$. 
We interpret dispersive lineshapes in the real part of the 1Q rephasing spectra as signatures of excitation-induced energy shifts and dephasing~\cite{Li2006, Yagafarov2020}. The $\pi$-shift difference between the polariton and $X_\text{A}$ lineshapes indicates that LP and MP$_1$ experience excitation-induced blueshifts, while $X_\text{A}$ undergoes a redshift. Dispersive lineshapes in 2D spectra caused by the blueshift of polariton states can arise due to Rabi contraction~\cite{Dunkelberger2016, Yagafarov2020, Allen2022} or polariton-polariton interactions~\cite{Sun2017}.
For systems with middle polaritons, such as the one described here, Rabi contraction should lead to a blueshift of the LP and MPs and a redshift of the UP. 
We estimate that, in this experiment, Rabi contraction can induce a maximum energy shift of $\sim2$\,meV~\cite{SM}.
Since the estimated energy shift is smaller than the homogeneous broadening of the diagonal features, Rabi contraction will result in dispersive lineshapes instead of off-diagonal peaks.
Additionally, Fieramosca et~al.~\cite{Fieramosca2019} reported that polariton-polariton interactions in (PEA)$_2$PbI$_4$, stemming from Coulombic effects amongst polaritons due to their excitonic constituents, induce blueshifts in the LP and MPs.
The redshift of $X_\text{A}$ can be related to exciton-exciton interactions or Coulomb mutual screening~\cite{Srimath2020Stochastic, gatjens2023manyexciton, xiao2023excitonexciton}.
Separating and quantifying the mechanisms that participate in these lineshapes is not straightforward, as various processes share similar signatures and have been modeled using different levels of theory~\cite{Li2006, Singh2016, Srimath2020Stochastic}. We have presented recently a velocity-gauge perspective to model coherent spectral lineshapes~\cite{bittner2025coherent}, which is ideally suited to model photoexcitations in crystalline semiconductors in the strong light-matter coupling regime; we will follow up with rigorous modelling in that perspective.  

The population-time evolution of the absolute 1Q rephasing~\cite{SM} reveals ultrafast incoherent population transfer from $X_\text{A}$ and MP$_\mathrm{1}$ to the LP in the (PEA)$_2$PbI$_4$ microcavity. 
Population transfer results in asymmetric cross-diagonal peaks, e.g.\ the ones marked with a square and a circle in Fig.~\ref{fig:1Q}(a), which indicate transfer pathways $X_\text{A} \rightarrow$ LP and MP$_\mathrm{1}$ $\rightarrow$ LP, respectively~\cite{Hao2016_CoherentIncoherent}. We note that the cross-peak within the square is apparently shifted with respect to the LP diagonal peak; however, we note that the spectral phase in the entire off-diagonal region is highly complex due to overlapping contributions. The apparent shift relaxes on a timescale corresponding to the relaxation of the spectral congestion~\cite{SM}. As population time progresses, the amplitudes of the cross-peaks increase relative to the other features in the two-dimensional map (Fig.~\ref{fig:1Q}(i)), supporting a downhill transfer from $X_\text{A}$ and MP$_\mathrm{1}$ to the LP occurring within a $\sim 100$\,fs timescale. We present the time-dependent intensities of diagonal- and cross-peaks as Supplemental Material~\cite{SM}.  
Similar scenarios have been observed in organic semiconductor microcavities~\cite{Mewes2020, Son2022} and addressed theoretically~\cite{Gallego2024}. 
In contrast, coherent population exchange (e.g. Rabi oscillations) between the lower and upper polariton govern the 1Q response of GaAs quantum-well microcavities~\cite{Pau2022_CoherentPolaritons}.

Importantly, the on-diagonal LP feature displays an apparent blueshift with increasing $t_\mathrm{pop}$, emphasized by tracking its contour (Fig~\ref{fig:1Q}(g)) and the spectrum of its diagonal cut (Fig.~\ref{fig:1Q})(h).  
The diagonal cuts at a series of population times (Fig.~\ref{fig:1Q}(h)), normalized at 2.340\,eV (the LP energy at $\vec{k}_\parallel = \Vec{0}$), show that the weight of the population of higher-energy larger-$\vec{k}_\parallel$ LP states grows in proportion to that of $\vec{k}_\parallel = \Vec{0}$ as $t_\mathrm{pop}$ increases.
Additionally, we observe hints of spectral features resembling cross-peaks between closely-spaced energetic states. The cross-peaks are marked with a diamond and a hexagon in Fig.~3. The hexagon feature suggests a back-transfer process from low to high energy states. 
Both observations indicate a population redistribution along the lower polariton branch towards states with higher energy and $\vec{k}_\parallel > \Vec{0}$, that is, a depletion of $\vec{k}_\parallel=\Vec{0}$ LPs within a $\sim 100$\,fs timescale caused by efficient polariton-polariton scattering.
The growth of the population weight at higher in-plane wavevector states can have a contribution from the shorter lifetime of LPs at $\vec{k}_\parallel=\Vec{0}$~\cite{SM}, due to their greater photonic component~\cite{Groenhof2019}. 
Nevertheless, we do not expect this contribution to be solely responsible for the depletion as the photonic component of the lower polariton is very similar at $\vec{k}_\parallel=\Vec{0}$ and slightly larger $\vec{k}_\parallel$. We rule out spectral diffusion effects in the 2D lineshape evolution because the exciton inhomogeneous bandwidth is only of similar order as the homogeneous linewidth~\cite{FelixEnhancedScreening}, and strong light-matter coupling filters within that inhomogeneous distribution; the polariton width is limited by the cavity lifetime. The blue shift in the lower polariton of $\sim 16$\,meV over 140\,fs in Fig.~\ref{fig:1Q}(g) is of similar magnitude to the width of the lower polariton at $\vec{k} = \vec{0}$ in Fig.~\ref{fig:cartoon_and_kspace}(c), which is not consistent with spectral diffusion in the 1Q rephasing dynamics of systems in a weak inhomogeneous limit. Furthermore, we do not observe spectral diffusion in the bare semiconductor over subpicosecond time windows~\cite{Srimath2020Stochastic}.  

To account for mode-mixing and spectral signatures observed in our experiments, we model inelastic scattering between polariton modes as a second-order virtual process mediated by a continuum of dark excitonic states. This mechanism resembles an anti-Stokes Raman process, wherein an initial lower-energy polariton state \( |1\rangle \) is scattered into a final higher-energy state \( |2\rangle \) via intermediate dark states \( \{ |n\rangle \} \) that are not directly coupled to the cavity modes. In our current formulation, we focused on the energy structure of intermediate dark excitonic states and treated the scattering amplitude as energy-resolved, implicitly integrating over momentum. This was motivated by the experimental geometry, where excitation and detection angles are fixed, and the dominant transitions correspond to small-angle scattering. That said, we also note that the presence of a broad continuum of dark excitonic states — spanning a wide range of energies and momenta — can make momentum conservation feasible even for large-angle scattering. These dark states can act as virtual intermediaries that absorb or supply the necessary momentum mismatch, particularly in systems with disorder, phonon coupling, or spatial inhomogeneity. The resulting transition amplitude is given by:
\[
T_{12}(\Delta\omega) = \int d\omega'\, \rho(\omega') \frac{g^2(\omega')\, n_{\mathrm{SS}}(\omega')}{\omega_1 + \Delta\omega - \omega' + i\eta},
\]
where \( \rho(\omega') \) is the dark-state density of states, \( g(\omega') \) the light--matter coupling, and \( n_{\mathrm{SS}}(\omega') \) the steady-state dark-state population. In thermal equilibrium, \( n_{\mathrm{SS}}(\omega') \propto e^{-\beta \omega'} \) and up-conversion processes are strongly suppressed. However, under cavity-driven nonequilibrium conditions, this population can become spectrally biased, enabling efficient up-scattering. We include a representative calculation of the inelastic scattering rate $k_{12} = |T_{12}(\Delta\omega)|^2$, shown in Fig.~S11 in the Supplemental Material~\cite{SM}. This provides a direct estimate of the transition rate as a function of energy detuning. The peak rate extracted from this calculation is on the order of $0.006$–$0.007$~fs$^{-1}$, consistent with the experimentally observed population redistribution occurring over $\sim$150~fs. We however note that our model is not fitted to the data, but rather provides a mechanistic framework that reproduces the qualitative features and order-of-magnitude rates observed experimentally. 

\begin{figure}[b!]
    \centering
    \includegraphics[width=8.6cm]{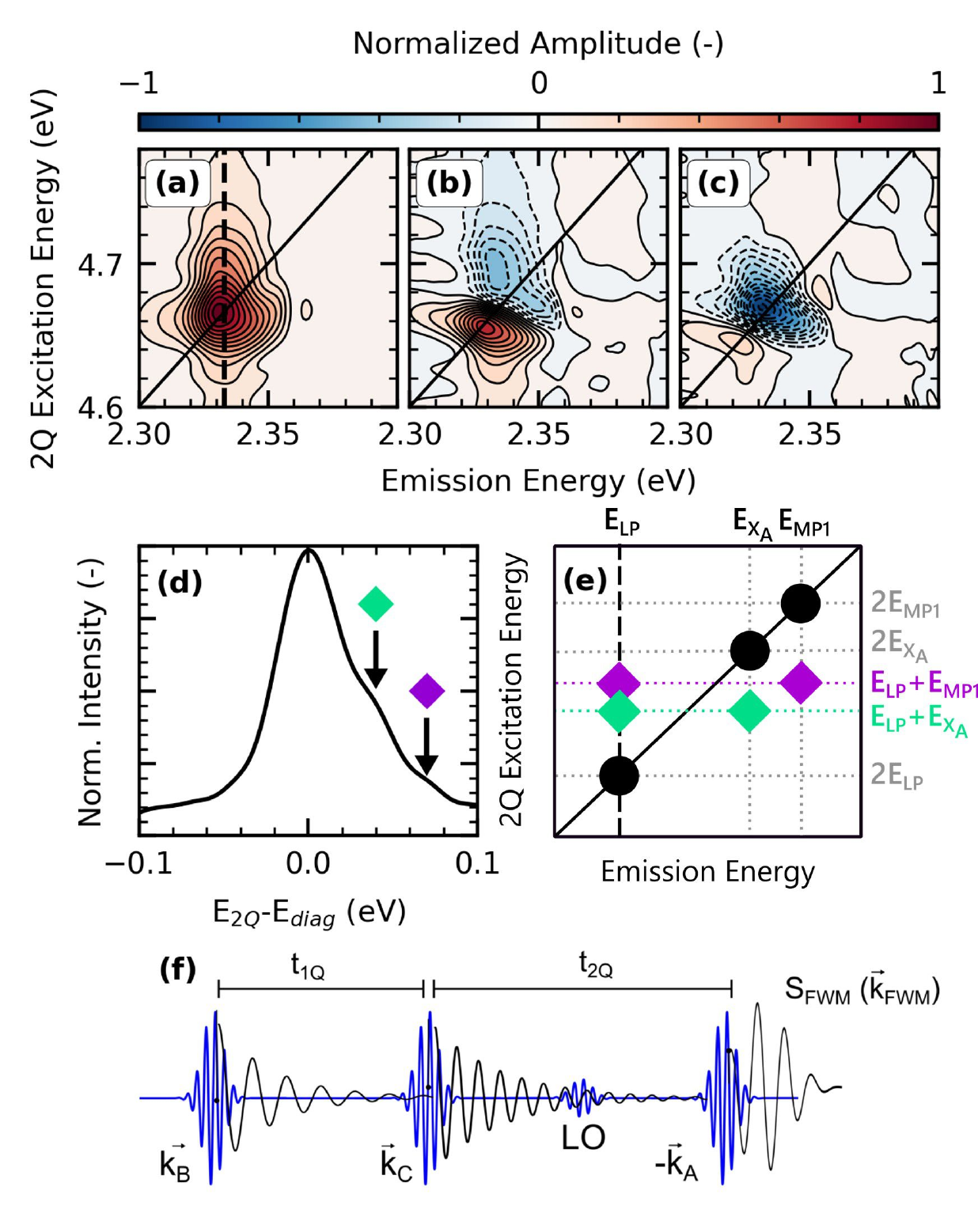}
    \caption{
    (a) Absolute, (b) real, and (c) imaginary components of the 2Q non-rephasing spectra measured at $t_\mathrm{1Q}=20$\,fs. 
    (d) A cut along the emission energy axis reveals two cross-peaks with energies 32\,meV (green diamond) and 65\,meV (purple diamond).
    (e) Schematic of a two-quantum non-rephasing spectrum of a system with LP--$X_\text{A}$ and LP--MP$_\mathrm{1}$ interactions, which manifest as cross-peaks with 2Q excitation energies of $E_\mathrm{LP}+E_{X_\mathrm{\text{A}}}$ and $E_\mathrm{LP}+E_\mathrm{MP}$, as indicated with green and purple diamonds respectively. 
    (f) Schematic representation of the pulse sequence employed to measure the 2Q nonrephasing spectra, using the same phase matching of the \textit{BoxCARS} geometry displayed in Fig.~\ref{fig:1Q}(h).}
    \label{fig:2Q}
\end{figure}

\section{Scattering between polaritons and the exciton reservoir}
Using two-dimensional two-quantum (2Q) spectroscopy~\cite{Yang2008, Wen_2013}, we elucidate two-particle correlations between excitons and polaritons~\cite{Stone2009_ExcitonExcitonCorrelations, Wen_2013, Autry2020} in the (PEA)$_2$PbI$_4$ microcavity.
For this, we utilize the pulse sequence shown in Fig.~\ref{fig:2Q}(f) and scan the time evolution of coherent states accessed via a two-step excitation ($t_\mathrm{2Q}$). 
In the absolute 2Q spectra (Fig.~\ref{fig:2Q}(a)), we observe a feature on the diagonal ($E_\mathrm{em}=2E_\mathrm{exc}$) corresponding to LP--LP self-correlations~\cite{Karaoskaj2010_Bi,Takemura2015_2Q2DFT}. 
We include the real and imaginary components of the two-quantum map in Figs.~\ref{fig:2Q}(b) and (c), correspondingly. 
The dispersive lineshape of the LP--LP feature in the real component of the spectrum indicates that multi-exciton interactions at the mean-field level are the principal contributors to the nonlinear signal, consistent with the prediction by Karaiskaj~\textit{et~al.}~\cite{Karaoskaj2010_Bi}. 
We see an asymmetry stretching above the diagonal ($E_\mathrm{em}=2E_\mathrm{exc}$) in the absolute (Fig~\ref{fig:2Q}(a)) and real component of the 2Q spectra (Fig.~\ref{fig:2Q}(b)).
We interpret this asymmetry as two closely spaced off-diagonal peaks, clearly resolved in a vertical cut at the emission energy of 2.335\,eV (dashed line in Fig.~\ref{fig:2Q}(a) and spectrum in Fig.~\ref{fig:2Q}(d)).
The difference between the energies of the off-diagonal and diagonal features ($E_\mathrm{2Q}-E_\mathrm{diag}$) are 32 and 65\,meV, which match with $E_\mathrm{X_\text{A}}-E_\mathrm{LP}$ and $E_\mathrm{MP_\mathrm{1}}-E_\mathrm{LP}$, respectively. 
This suggests that the off-diagonal features correspond to LP--$X_\text{A}$ and LP--MP$_\mathrm{1}$ two-particle states, as indicated with green and purple diamonds, correspondingly, in Figs.~\ref{fig:2Q}(d-e)~\cite{Takemura2015_2Q2DFT}.
Generally, as indicated in Fig.~\ref{fig:2Q}(e), at emission energies $E_{X_\text{A}}$ and $E_{MP_\mathrm{1}}$, one expects self-interactions from $X_\text{A}$ and $MP_\mathrm{1}$ (in the diagonal) and cross-correlations between LP--$X_\text{A}$ and LP--$MP_\mathrm{1}$ (asymmetries below the diagonal).
There are many reasons that asymmetric
peak amplitudes can occur in 2D coherent spectra, for example, related to interference between
different pathways; we do not address this point here. 
Concisely, we demonstrate that both excitons ($X_\text{A}$) and middle polaritons (MP$_\mathrm{1}$) act as scattering reservoirs for lower polariton states and play an integral part in the polariton dynamics of two-dimensional metal-halide microcavities.

\section{Conclusion}

In summary, beyond 1\,ps, nonlinear dynamics are largely governed by the dynamics of the reservoir population. Multiparticle interaction-driven annihilation is evident in the emission from the lower polariton state. On $< 150$-fs timescales, we observe ultrafast population transfer from resonantly pumped higher-energy polariton states and reservoir populations to the $\vec{k} = \Vec{0}$ state of the LP dispersion. Concurrently, LP depletion from $\vec{k} = \vec{0}$ to $|\vec{k}| > \vec{0}$ suggests many-body interactions driving dephasing toward the exciton reservoir~\cite{Takemura2016, Takemura2016Dephasing}. We also present direct evidence of many-body correlations between LP and $X_{A}$, as well as MP$_\mathrm{1}$, identifying them as Coulomb scattering baths coupled to the LP. While mean-field models capture the general polariton dynamics, they may overlook fluctuation-driven effects arising from these scattering pathways, which have a measurable impact~\cite{Wouters, ERB2012a,ERB2012b,Svitlana2015}. Our findings on population transfer timescales and energy correlations offer a strong experimental foundation to refine these theoretical frameworks and reveal material-specific mechanisms underlying polariton condensation.

\begin{acknowledgments}
VQC, ERG, MGD, CARP, JPCB, and CSA acknowledge support from the National Science Foundation Science and Technology Center (STC) for Integration of Modern Optoelectronic Materials on Demand (IMOD) under award number DMR-2019444.
VQC thanks the Georgia Tech Quantum Alliance and SPIE Optics and Photonics Education Scholarships for graduate student support.
CARP acknowledges support from the National Aeronautics and Space Administration (Award\#80NSSC19M0201).
ERB was funded by the National Science Foundation (CHE-2404788) and the Robert A.\ Welch Foundation (E-1337). 
ARSK acknowledges start-up funds from Wake Forest University, funding from the Center for Functional Materials, and funding from the Office of Research and Sponsored Programs at Wake Forest University.
HL and CSA acknowledge funding from the Government of Canada (Canada Excellence Research Chair CERC-2022-00055). CSA acknowledges support from the Institut Courtois, Facult\'e des arts et des sciences, Universit\'e de Montr\'eal (Chaire de recherche du directeur de l'Institut Courtois).
\end{acknowledgments}

\section{Data availability statement}
The data supporting the findings of this article are openly available at~\cite{Quiros_data_2025}.

\providecommand{\noopsort}[1]{}\providecommand{\singleletter}[1]{#1}%
%


\clearpage
\newpage
\renewcommand{\thepage}{S\arabic{page}}
\renewcommand{\thefigure}{S\arabic{figure}}
\renewcommand{\thesection}{S\arabic{section}}
\renewcommand{\thesubsection}{\alph{subsection}}
\setcounter{page}{1} 
\setcounter{figure}{0} 
\setcounter{section}{0} 
\begin{center}
    \noindent\large{\textbf{Supplemental Material: Resolving exciton and polariton multi-particle correlations in an optical microcavity in the strong coupling regime}} \\
  \vspace{0.5cm} 

  \normalsize

\noindent{Victoria~Quir\'os-Cordero, Esteban~Rojas-Gatjens, Martin Gomez-Dominguez, Hao~Li, Carlo~A.~R.~Perini, Natalie~Stingelin, Juan-Pablo~Correa-Baena, Eric~R.~Bittner, Ajay~Ram~Srimath~Kandada, Carlos~Silva-Acu\~na}\\
\end{center}


\section*{Materials and Methods}

\subsubsection*{(PEA)$_2$PbI$_4$ microcavity fabrication}
The microcavity comprises a bottom quarter-wavelength distributed Bragg reflector (DBR) with 10.5 pairs of TiO$_2$/SiO$_2$, a 60\,nm (PEA)$_2$PbI$_4$ spin-coated film, a 125\,nm poly(methyl methacrylate) (PMMA) spacer, and a top 40\,nm silver mirror.
The DBR has a stopband centered at 520\,nm and was purchased from SpectrumThinFilms. 
For the microcavity preparation, first, we cleaned the DBR in sequential ultrasonic baths of acetone and IPA for 15 min each. 
Then, we dried it with nitrogen and treated it with UV-ozone for 15 min.
The two-dimensional perovskite precursor solutions were prepared by dissolving PbI$_2$ (purity $>$ 99.99\.\%) and phenethylammonium iodide (purity $>$ 99.99\,\%) in N,N-Dimethylformamide (purity $>99.98$\%) at a 0.13\,M concentration.
The two-dimensional perovskite films were deposited by dropping 80\,$\mu$L of precursor solution on the 1\,in$^{2}$ clean DBRs before spin coating them at 6000 rpm for 30\,s with an acceleration of 6000\,rpm/s.
Immediately after, the films were annealed for 10\,min at 100\,$^\circ$C.
The PMMA solution was prepared by dissolving 30\,mg of PMMA (M$_\mathrm{w}$ $\sim$15,000\,g/mol) in 1\,mL of toluene (purity $>$ 99.98\,\%).
The PMMA solution was then deposited by spin coating.
80\,$\mu$L of the solution was dropped on the two-dimensional perovskite layer and spin-coated employing a 2-step process, with an initial spreading step at 100 rpm accelerated at 1000\,rpm/s for 10\,s, followed by a 6000\,rpm step accelerated at 6000\,rpm/s.
The stack was heated for 5 minutes at 60\,$^\circ$C to dry the PMMA film. 
Physical vapor deposition (PVD) was used for the top silver mirror; silver pellets (purity $>$ 99.999\,\%) were thermally evaporated at a rate of 0.5\,angstrom/s to a final thickness of 42\,nm.

\subsubsection*{Fourier imaging}\label{FM}

Using a home-built Fourier microscope, we imaged the energy dispersion of the reflectance and photoluminescence.
The microscope employs a Zeiss LD EC Epiplan Neofluar 100X infinity-corrected objective (NA = 0.75), an Acton SpectraPro 300i spectrometer, and an Andor Newton EM camera. 
For reflectance and photoluminescence measurements, we use a ThorLabs SLS201L broadband light source and the output of an optical parametric amplifier (ORPHEUS, Light Conversion) at 470\,nm pumped by a PHAROS laser (Model PH1-20-0200-02-10, Light Conversion), respectively. 

\subsubsection*{Excitation correlation photoluminescence (ECPL)}

The excitation correlation photoluminescence setup used as a laser source a PHAROS (Model PH1-20-0200-02-10, Light Conversion) with an output operating at a repetition rate of 100\,kHz. 
A portion of the laser beam was sent into a collinear optical parametric amplifier (ORPHEUS, Light Conversion) tuned for an output of 470\,nm. 
The pulse train was then split 50/50 by a beam splitter cube, and one of the pulse trains was directed to a motorized linear stage (LTS300, Thorlabs), to impart a time delay between the two pulses. 
Both pulse trains were amplitude-modulated with a chopper at distinct frequencies (522 and 700\,Hz).
The pulse trains were then recombined and focused onto the sample using the same setup described in the Fourier Microscopy section. 
A flip mirror allowed us to alternate between the camera to measure the photoluminescence energy dispersion and an avalanche photodiode (APD440A, Thorlabs) for the ECPL experiment. 
The total integrated response was demodulated using a lock-in amplifier (HF2LI, Zurich Instruments), at the fundamental and sum of the modulation frequencies to isolate the linear and nonlinear components of the photoluminescence. 

\subsubsection*{Two-dimensional coherent spectroscopy}
A detailed description of the setup implemented can be found in our previous work~\cite{Thouin2018Bi, FelixEnhancedScreening}. 
Briefly, a portion of the 1030-nm, 220-fs output of a PHAROS laser (Model PH1-20-0200-02-10, Light Conversion), operating at a 100-kHz repetition rate, was directed into a home-built third-harmonic-pumped non-collinear optical parametric amplifier. 
A beam geometry consisting of four pulse trains in a \textit{BoxCARS} geometry was generated using a diffractive optical element. The four pulse trains are then sent to a home-built pulse shaper which compresses the pulses individually using a second harmonic generation chirp scan.
Additionally, the pulse shaper applies a delay between the pulses and performs the phase cycle during the measurement.
The resulting pulse duration was 20\,fs full-width at half-maximum, measured by second-harmonic generation cross-frequency-resolved optical gating (SHG-XFROG), as characterized in Supplementary Information. 
All measurements were carried out in a vibration-free closed-cycle cryostat (Montana Instruments).  The spot size was $95\pm5\,\mu$m.
The spectra were collected by pumping with the same laser spectrum and an incident fluence of 0.7\,$\mu$J\,cm$^{-2}$ per pulse. 
The pulses excite the sample at an angle of 2.3\,$^\circ$, which corresponds to $|\vec{k}_\parallel | =0.92\,\mu\mathrm{m}^{-1}$.

\section*{Microcavity design}

The energy of the microcavity optical mode was designed such that the lower polariton has good energetic overlap with the maximum emission of (PEA)$_2$PbI4$_4$ at small in-plane wavevectors.
This design should facilitate the population of the lower polariton $\vec{k}_{\parallel} =\vec{0}$ state via radiative pumping~\cite{RadPumpCarlosLidzey, Deshmukh:23, Laitz2023} and the observation of polariton-polariton correlations.

We used the following expressions to calculate the microcavity quality factor $Q$, the optical mode lifetime $\tau_\text{opt}$, and the lower polariton lifetime $\tau_\text{LP}$: 
\begin{equation}
    Q\,=\,E_\text{opt}(\vec{k}_{\parallel} =\vec{0})/\Delta_{\text{FWHM}}[E_\text{opt}(\vec{k}_{\parallel} =\vec{0})]
\end{equation}
\begin{equation}
\tau_\text{opt} = 2Q/\omega_\text{p}(\vec{k}_{\parallel} =\vec{0})
\end{equation}
\begin{equation}
\tau_\text{LP} = \left[|X_\mathrm{opt}|^2/\tau_\mathrm{opt} + \sum_{i=\text{A},\text{B},\text{C}} |X_i|^2/\tau_\mathrm{X_i}\right]^{-1}
\end{equation}

Where $E_\text{opt}(\vec{k}_{\parallel} =\vec{0})$ is the optical mode energy at normal incidence according to the transfer matrix simulation detailed in the next section, $|X_\mathrm{opt}|^2$ and $|X_i|^2$ are the Hopfield coefficients of the lower polariton at $\vec{k}_\parallel=\vec{0}$ (see the strong light-matter coupling Hamiltonian section), and $\tau_\mathrm{X_i}$ is the lifetime of the excitons ($i=\text{A}, \text{B}, \text{C}$) reported in Ref.~\citenum{Straus2016} at 15\,K.

\section*{Transfer matrix model simulation}

We simulated the energy dispersion of the microcavity optical mode using a transfer matrix model (TMM). 
We followed the TMM equations \eqref{eq:SI_TMMfirst} -- \eqref{eq:SI_TMMlast} described below and used the experimentally determined thicknesses and complex energy-dependent refractive indices of every layer in the microcavity stack, except for the (PEA)$_2$PbI$_4$.
To simulate the dispersion of the optical mode, instead of that of the polariton branches via TMM, we must consider the refractive index of (PEA)$_2$PbI$_4$ to be static, real, and equal 2.61, which is the real refractive index of the material at the target optical mode energy of 2.436 eV.
The simulated dispersion of the optical mode is shown in Fig. \ref{fig:SI_OptMode}.
We obtained equation \eqref{eq:OptDispersion} describing the dispersion as a function of in-plane wavevector by fitting a polynomial to the reflectance minimum (red dashed line in Fig.~\ref{fig:SI_OptMode}) 
\begin{equation}
    E_{\mathrm{opt}}(\vec{k}_{\parallel}) = \hbar \omega_\text{P}(\vec{k}_{\parallel}) = 2.436 + 9.679 \times 10^{-4} \vec{k}_{\parallel}^2 \,\mathrm{eV}.
\label{eq:OptDispersion}
\end{equation}
Next, we summarize the main equations of the transfer matrix model. More details can be found in Ref.~{\cite{TMM}}. The optical response of a thin film can be fully described by a transfer matrix that summarizes the continuity conditions of the electric and magnetic field components that are tangential to the film's interfaces.
The angle-dependent transfer matrix of a single film is given by:
\begin{gather}
T_{\mathrm{film},\,s-\mathrm{pol}} 
= \begin{pmatrix}
\cos(\delta_\mathrm{film}) & i \sin(\delta_\mathrm{film})/[n_\mathrm{film}\cos(\theta_\mathrm{film})] \\
i n_\mathrm{film} \sin(\delta_\mathrm{film})\cos(\theta_\mathrm{film})& \cos(\delta_\mathrm{film})\\
\label{eq:SI_TMMfirst}
\end{pmatrix},
\end{gather}

\begin{gather}
T_{\mathrm{film},\,p-\mathrm{pol}} 
= \begin{pmatrix}
\cos(\delta_\mathrm{film}) & i \sin(\delta_\mathrm{film})\cos(\theta_\mathrm{film})/n_\mathrm{film} \\
i n_\mathrm{film}\sin(\delta_\mathrm{film})/\cos(\theta_\mathrm{film})& \cos(\delta_\mathrm{film}) \\
\end{pmatrix},
\end{gather}
\begin{equation}
\delta_\mathrm{film} = \frac{2\pi n_\mathrm{film} d_\mathrm{film}}{\lambda} \cos(\theta_\mathrm{film}),
\end{equation}
where $n_\mathrm{film}$ corresponds to the complex energy-dependent refractive index, $\theta_\mathrm{film}$ is the angle of propagation of light as given by Snell’s law, and $d_\mathrm{film}$ the film thickness.
This result can be extended to an assembly of \textit{q} thin films, where the total transfer matrix is the product of the individual matrices,  $T_\mathrm{assembly} = \Pi^q_\mathrm{film}T_\mathrm{film}$. The left-most matrix corresponds to the first film light traverses.

The reflectance (R) of a thin-film assembly can then be obtained from its transfer matrix as follows: 
\begin{gather}
\begin{pmatrix}
E_\mathrm{in}/E_\mathrm{out} \\
H_\mathrm{in}/H_\mathrm{out} 
\end{pmatrix} 
= 
\begin{pmatrix}
B \\
C 
\end{pmatrix} 
= T_\mathrm{assembly}
\begin{pmatrix}
1 \\
n_s
\end{pmatrix},
\end{gather}

\begin{equation}
    R = \left(\frac{n_0B-C}{n_0B+C}\right)\left(\frac{n_0B-C}{n_0B+C}\right)^*,
\label{eq:SI_TMMlast}
\end{equation}
where $n_0$ and $n_s$ correspond to the refractive indices of the incident and output media (typically air and the substrate, correspondingly).

\section*{Strong light-matter coupling Hamiltonian}

The following multiple-exciton Hamiltonian, based on a Jaynes-Cummings model extended to three excitonic states and using the rotating wave approximation, describes the polariton branches observed in the microcavity energy dispersion. Notably, to reproduce the energy dispersion of the polariton branches observed experimentally, at least three distinct excitonic states are required.  

\begin{equation}
    \hat{H}(\vec{k}_\parallel) = \sum_{i=\text{A}, \text{B}, \text{C}}\hbar\omega_i b^\dagger_ib_i
    +\hbar\omega_\text{p}(\vec{k}_\parallel) a^\dagger_\text{p} a_\text{p} +\sum_{i=\text{A}, \text{B}, \text{C}} g_i(a^\dagger_\text{p}b_i + a_\text{p} b^\dagger_i). \label{eq:Hamil}
\end{equation}

The first term represents the creation and annihilation of excitons ($i = \text{A,}\:\text{B,}\:\text{C}$) with energy $\hbar\omega_i$ and excitonic annihilation operator $b_i$. The second term describes the insertion and removal of cavity-mode photons as a function of in-plane wavevector whose energy is a function of in-plane wavevector $\hbar\omega_\text{p}(\vec{k}_\parallel)$ and with photon annihilation operator $a_\text{p}$.
The third term describes strong exciton-photon coupling, where $g_i$ is the coupling constant between the $i$-th exciton and the cavity-mode photons. 

We diagonalized the Hamiltonian using the numerical package Qutip~{\cite{Johansson2013_Qutip}} and leveraged the dispersion bands to estimate coupling strengths $g_\text{A,B,C} = {80, 80, 90}$\,meV. 
During this process, we fixed the energy of the excitonic transitions (2.370, 2.404, and 2.445\,eV) and the microcavity dispersion (calculated using the transfer matrix model as described above) while changing the coupling strengths. 
We emphasize that since we have three coupling strengths as variables, various combinations of them reproduce the polariton branches observed with good accuracy, and hence these coupling strengths should not be taken as definitive.
However, we note that the couplings are of the order of $\approx80$ -- 100\,meV and that at least three excitons are required to reproduce all polariton band energies.
The simulated polariton eigenstates are shown in Fig.~1 as solid white lines overlaid to the experimental energy dispersion at 5\,K. 
Additionally in Fig.~\ref{fig:HCoeffs}, we show the Hopfield coefficients for the polariton states that diagonalize the Hamiltonian. 
These coefficients provide information on the photonic and excitonic components that constitute each state.

\section*{Reflectance energy dispersion}
\subsubsection*{Observation of four polariton branches} We distinguish four polariton branches in the experimental energy dispersion of the microcavity: an upper (UP), two middle (MP1 and MP2), and a lower polariton. We emphasize this in Fig.~\ref{fig:ReflCuts} by displaying reflectance spectra at a series of in-plane wavevectors. Each polariton branch corresponds to a low-reflectivity peak.

\subsubsection*{Factors influencing the reflectance of polariton branches.} 
\begin{figure}
    \centering
    \includegraphics[width=\textwidth]{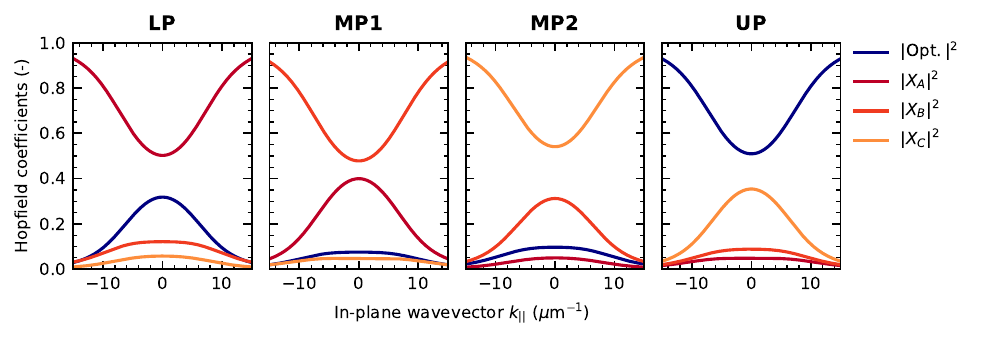}
    \caption{Hopfield coefficients of the polariton eigenstates. They were calculated by numerically diagonalizing the Hamiltonian (1) of the main text using the numerical package, Qutip. The simulation details are discussed in the text above.}
    \label{fig:HCoeffs}
\end{figure}

\begin{figure}
    \centering
    \includegraphics[width=0.6\textwidth]{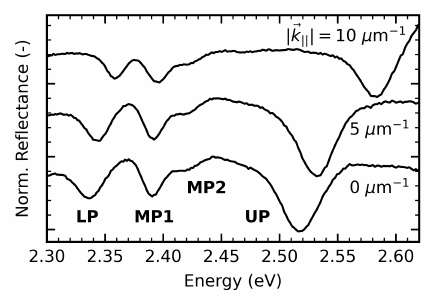}
    \caption{Reflectance spectra of the (PEA)$_2$PbI$_4$ microcavity at in-plane wavevectors of 0, 5, and 10 $\mu$m$^{-1}$ showing four polariton branches.}
    \label{fig:ReflCuts}
\end{figure}

\begin{figure}
    \centering
    \includegraphics[width=0.6\textwidth]{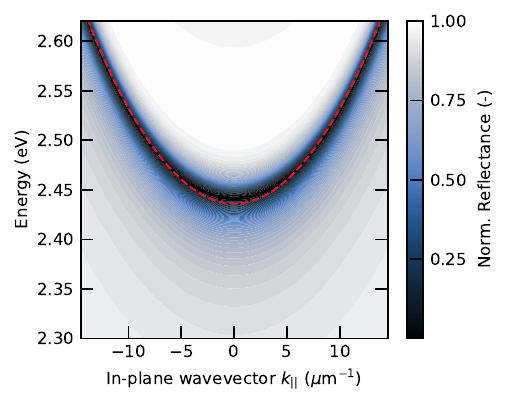}
    \caption{Transfer matrix model of the microcavity optical mode. We used the TMM method described above and fitted to the simulation a quadratic polynomial (red dashed line) to get a mathematical expression for the dispersion.}
    \label{fig:SI_OptMode}
\end{figure}

The difference in reflectance among polariton states is likely because of their distinct $X_\text{A}$, $X_\text{B}$, and $X_\text{C}$ fractions (\eqref{fig:HCoeffs}), since these excitonic constituents have different oscillator strengths.
This is evidenced in the absorption spectra of (PEA)$_2$PbI$_4$ (left panel, Fig. 2 in the main text).
In addition, variations in the overlap of the polariton states with the exciton reservoirs can also lead to differences in reflectance. It has been reported previously that, in the most exacerbated case, this overlap causes ``a polariton breakdown'' in which the reflectance of the state vanishes~\cite{UNAMBreakdown}.

\section*{Photoluminescence-detected techniques}

\subsubsection*{Linear measurements} 
\begin{figure}
    \centering
    \includegraphics[width=\textwidth]{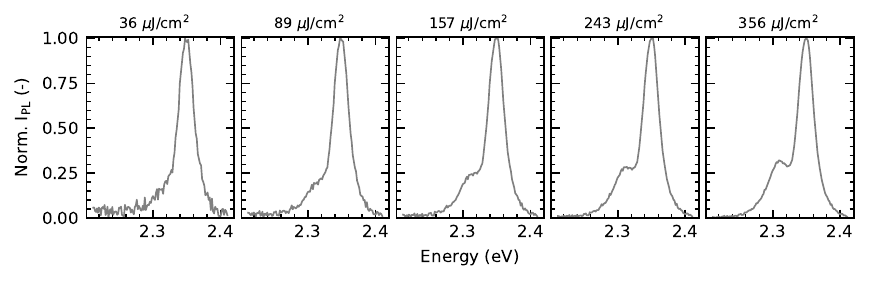}
    \caption{Photoluminescence spectra of a 60 nm (PEA)$_2$PbI$_4$ as a function of fluence. This data was taken under a non-resonant excitation (470 nm, 220 fs) with varying fluence and at a temperature of 5\,K.}
    \label{fig:SI_PEA2PbI4_PL}
\end{figure}

We show the fluence-dependent photoluminescence (PL) spectra of the bare (PEA)$_2$PbI$_4$ film as a function of fluence (Fig.~\ref{fig:SI_PEA2PbI4_PL}), in which we observe a secondary emission assigned in the literature to a dark exciton, a biexciton state, and the overlap between both~\cite{Kondo1998_Biexciton, polimeno2020observation,Laitz2023}.

\begin{figure}
    \centering
    \includegraphics[width=\textwidth]{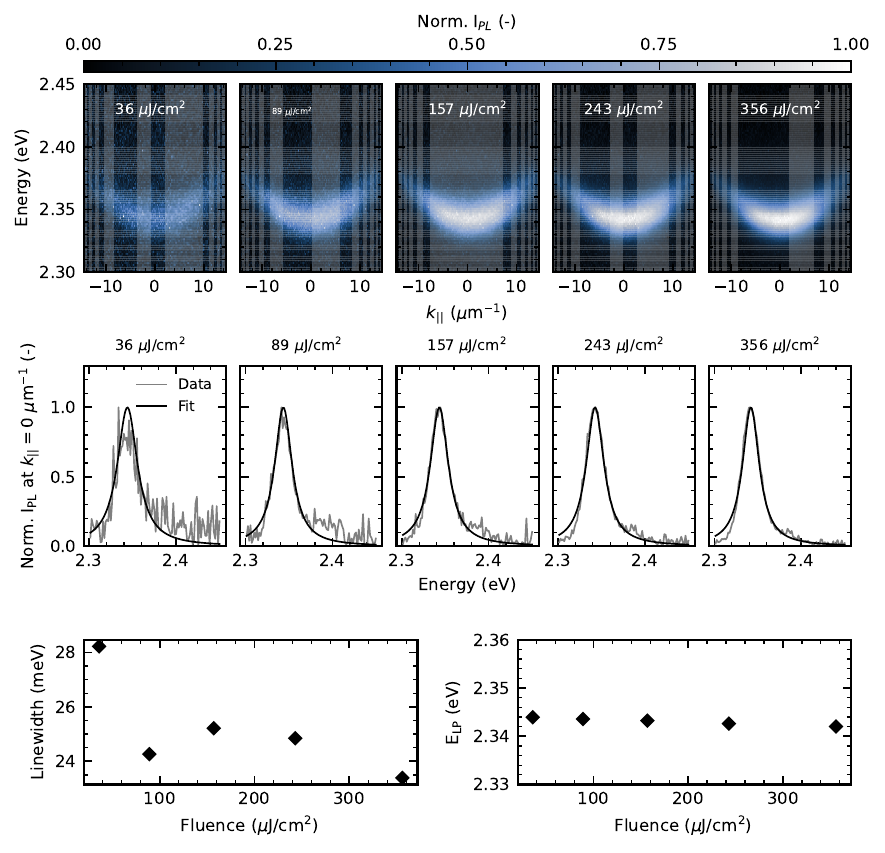}
    \caption{Photoluminescence energy dispersion of the lower polariton as a function of fluence. We show the PL spectra at $\vec{k}_\parallel = 0$, linewidth, and energy  of the (PEA)$_2$PbI$_4$ microcavity as a function of fluence measured under non-resonant excitation (470 nm, 220 fs).}
    \label{fig:MC_PL1}
\end{figure}

Additionally, we include the PL energy dispersion of the lower polariton and a spectral cut at $\vec{k}_\parallel=\vec{0}$ (Fig.~\ref{fig:MC_PL1}, top panel) as a function of fluence. 
These measurements were taken under non-resonant pumping (470\,nm, 220\,fs), using the same laser pulse employed in the ECPL experiments, and detecting with an Andor Newton EM camera.
We also display the linewidth and the maximum peak energy of the lower polariton emission as a function of fluence, extracted from Lorentzian fits to the spectra (Fig.~\ref{fig:MC_PL1}, bottom panel).

\subsubsection*{Nonlinear measurements: Excitation correlation photoluminescence (ECPL) spectroscopy} Briefly, ECPL is a technique that excites the sample with two pulses, one of which is temporally delayed from the other.
The pulses are amplitude-modulated at two distinct frequencies ($\Omega_1$ and $\Omega_2$).
Then, we detect the time-integrated photoluminescence and isolate its nonlinear component by demodulating at the sum frequency ($\Omega_1+\Omega_2$) via lock-in detection. 
The time resolution arises from the delay between the two pulses.

\subsubsection*{Modeling bimolecular annihilation in a semiconductor}
A simple kinetic model for exciton dynamics, discussed in more detail in Ref.~\cite{Rojas2023ECPL}, accurately describes the ECPL traces of (PEA)$_2$PbI$_4$ films. 
This model includes a monomolecular recombination term with a $\gamma$ rate and an exciton-exciton annihilation term with a parameter $\beta$.
The differential equation and its corresponding solution are:
\begin{equation}
\begin{split}
    \frac{dn}{dt} &= -\gamma n - \beta n^2 \\
    n(t) &= \frac{n_0\gamma/\beta}{(n_0+\gamma/\beta)\exp(\gamma t)-n_0}.
\end{split}
\end{equation}
From the time-dependent exciton population $n(t)$, we can obtain the total photoluminescence after the interaction with two identical laser pulses, assumed to be delta functions and, therefore, considered as initial conditions. The total photoluminescence detected is:
\begin{equation}
    I_\text{total\,PL} \propto \int_0^\tau n_1(t) dt+\int_\tau^\infty n_2(t-\tau)dt.
\end{equation}
The first integral considers the time when only one pulse arrived at the sample and the second integral considers the time after the two incident pulses.
The initial conditions correspond to $n_1(0) = n_0$ and $n_2(0) = n_1(\tau)+ n_0$, where $n_1(\tau)$ is the residual population due to the sample's interaction with the first pulse.
Then we subtract the single pulse contributions, $I = 2\int_0^\infty n(t)dt = 2n_0$, to obtain the nonlinear photoluminescence:
\begin{equation}
    \Delta I_\text{PL} (\tau) \propto \ln \left(1-\frac{\alpha^2 \exp(-\gamma \tau)}{(1+\alpha)^2}\right) \: ; \:\alpha=\frac{n_0\beta}{\gamma}.
    \label{Eq:dPL2}
\end{equation}
Note that although both time-resolved photoluminescence and ECPL follow the same population, their decay functions are different and we cannot compare their decay traces directly.

\subsubsection*{Modeling the lower-polariton picosecond dynamics}
We rationalize the lower-polariton picosecond dynamics by expanding the kinetic model for the semiconductor exciton dynamics.
Following Ref.~\cite{Mazza2013}, we include a differential equation for the time evolution of the lower polariton population $p(t)$ that is coupled to the time-dependent exciton-reservoir population $n(t)$.
This additional equation includes the radiative decay of the LP with a rate of $\gamma_\text{LP}$, bimolecular annihilation between the exciton reservoir and the lower polariton ---which depends on $|X_\text{A}|^2$, the $X_\text{A}$ fraction of the LP, and is described by the parameter $\beta_\text{LP}$--- and a transfer rate $W_{\text{exc}\rightarrow\text{LP}}$ from the reservoir to the lower polariton.
\begin{equation}
\begin{split}
    \frac{dn}{dt} &= -(\gamma + W_{\text{exc}\rightarrow\text{LP}}) n - \beta n^2\\
    \frac{dp}{dp} &= -\gamma_\text{LP}p - \beta_\text{LP}|X_\text{A}|^2np + W_{\text{exc}\rightarrow\text{LP}}n.
\end{split}
\end{equation}
These coupled differential equations do not have an analytical solution. They must be solved numerically to find $p(t)$. Then, the total photoluminescence of the microcavity is given by:
\begin{equation}
    I_\text{total\,PL} \propto \int_0^\tau p_1(t) dt+\int_\tau^\infty p_2(t-\tau)dt.
\end{equation}
With initial conditions $p_1(0) = 0$ and $p_2(0) = p_1(\tau)$, where $p_1(\tau)$ is the residual LP population due to the sample's interaction with the first pulse.
The nonlinear photoluminescence of the LP is $\Delta I_\text{PL}=I_\text{total\,PL}-I$, where $I$ corresponds to the lower-polariton PL intensity originating from two single pulses and can be expressed as $I = 2\int_0^\infty p(t)dt$.
Bimolecular annihilation between lower polaritons and excitons plays a minor role since polaritons are short-lived compared to excitons.
From this model, it is clear that exciton bimolecular annihilation in the reservoir is the main nonlinearity quenching the total photoluminescence of lower polaritons in picosecond timescales and leading to a negative ECPL signal. 

\section*{Two-dimensional coherent spectroscopy}
\subsubsection*{Characterizing pulse duration} 
\begin{figure}
    \centering   
    \includegraphics[width=0.6\textwidth]{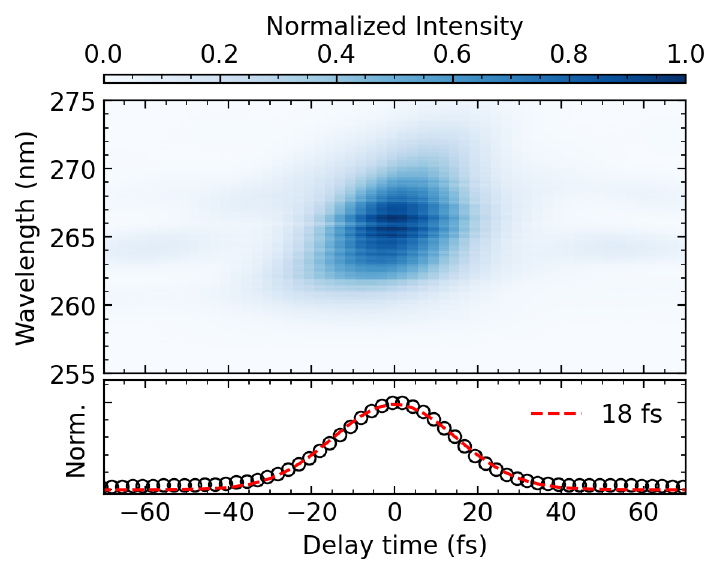}
    \caption{X-Frog characterization of the pulse. The X-frog is performed by placing a thin BBO at the sample position. The durations is extracted by fitting a Gaussian function.}
    \label{fig:SI_FROG}
\end{figure}

The second-harmonic generation cross-frequency-resolved optical gating (SHG-XFROG) of the pulse employed is shown in Fig. \ref{fig:SI_FROG}, which indicates that our pulse duration is 20\,fs, full-width half-maximum. 

\subsubsection*{Population-time series of the microcavity's one-quantum (1Q) rephasing response} 
\begin{figure}
    \centering
    \includegraphics[width=\textwidth]{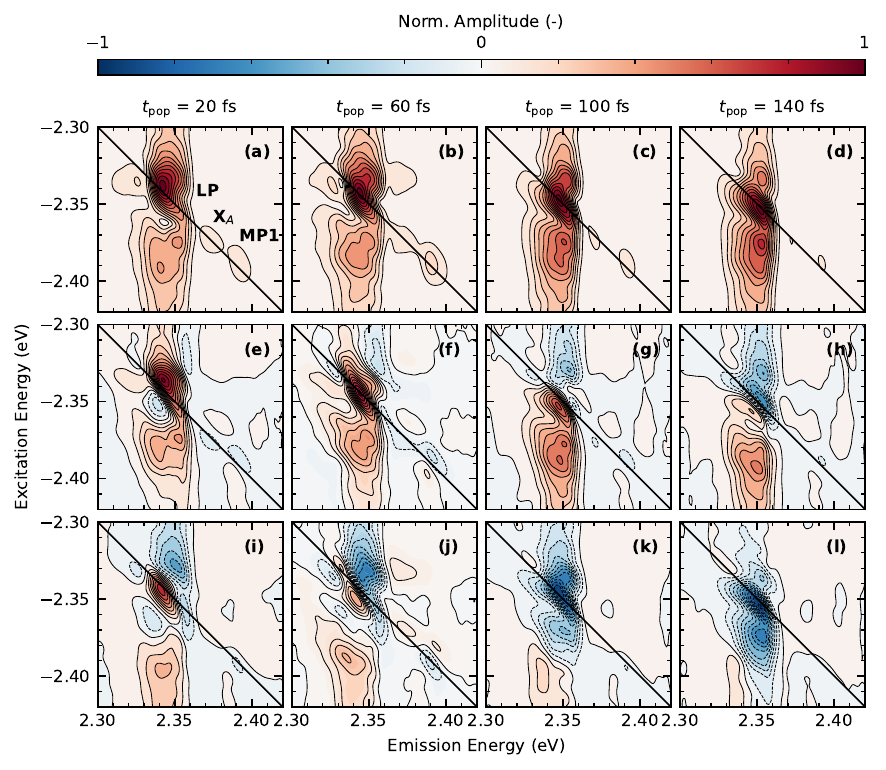}
    \caption{Ultrafast energy transfer from $X_\text{A}$ and MP1 to the lower polariton. We show the evolution of the rephasing 2D coherent spectra as a function of population time ($t_\mathrm{pop}$) for the (PEA)$_2$PbI$_4$ microcavity measured at 10\,K. The subfigures (a-d) correspond to the absolute components, (e-h) correspond to the real component and (i-l) correspond to the imaginary component.} 
    \label{fig:1QSI}
\end{figure}

The 1Q rephasing spectra of the (PEA)$_2$PbI$_4$ microcavity at various population times ($t_\text{pop}$) can be found in Fig. \ref{fig:1QSI}. Its absolute component is displayed in Fig. \ref{fig:1QSI}(a-d) (top row), while its real and imaginary components are included in Fig. \ref{fig:1QSI}(e-h) (middle row) and Fig. \ref{fig:1QSI}(i-l) (bottom row), respectively. We summarized the main features of the absolute 1Q rephasing spectra evolving with $t_\text{pop}$ in Fig. 5 of the main text. The vertical and diagonal cuts and the lower polariton contour were extracted from the data plotted here.

\subsubsection*{Contributions to the 1Q and 2Q signals} In Fig.~\ref{fig:FeynmanSI}, we show the pathways contributing to the 1Q rephasing (top) and 2Q non-rephasing spectra (bottom) included in the main text.
Contributions to 1Q rephasing spectra include ground-state bleach (GSB), stimulated emission (SE), and excited-state absorption (ESA), which involve both singly and doubly excited-state pathways. 
The 2Q spectrum solely probes doubly excited-state pathways, resolving the double excitation manifold without contributions from single excited states.


When considering the bosonic nature of excitons, ESA pathways can result in dispersive lineshapes and linewidth broadening in the 1Q rephasing spectra, as the stabilization energy of ij doubly excited states ---in this work, i,j = LP, MP$_1$, X$_\text{A}$, ...--- typically leads to a redshifted feature that overlaps with the ground-state bleach.
This has been discussed by Singh\textit{et al.}~\cite{Singh2016}. 
However, dispersive-like 1Q rephasing spectra are usually assigned to energy shifts that result from a convolution of different mechanisms, such as exciton-exciton interactions, background exciton scattering, and Rabi contraction.
Different physical models have been presented to account for dispersive-like spectra in two-dimensional spectroscopy.
This dispersive lineshape is also explained by employing phenomenological models based on modified optical Bloch equations~\cite{Li2006} and a stochastic theory derived by some of the authors in previous works~\cite{Srimath2020Stochastic}.
Finally, in the case of polaritons, Rabi contraction (a decrease in the Rabi splitting due to exciton photo-bleaching induced by the excitation) also results in dispersive lineshapes.
The distinct physical processes leading to the dispersive lineshapes and the many models that describe them make the simulation theoretically demanding and outside the scope of this work. 
Additionally, we note that, in reality, one never has one, two, or three excitations in these systems. A proper many-body description demands that one work in the large-N limit and that excitations are taken from an ensemble of (at best) weakly interacting 
bosons. Recent theoretical work by P\'erez~\cite{PerezSanchez2023} has explored this limit. 

\subsubsection*{Rabi contraction} We estimated the Rabi contraction effect roughly, as described by equation~\eqref{EqA}.
    \begin{equation}
    g_i = 2\Omega_i = g_0\sqrt{N_i}.
    \end{equation}
    \begin{equation}
     \Delta = 2\Omega_1-2\Omega_2 = g_0\sqrt{N_1}-g_0\sqrt{N_2}.  
    \end{equation}    
    \begin{equation}
        \Delta = g_1\left(1-\sqrt{\frac{N_2}{N_1}}\right),
        \label{EqA}
    \end{equation}
    where $N_1$ is the initial number of excitonic transitions and $N_2$ is the remaining number of transitions after excitation. We estimate $N_1$ using the exciton's Bohr radius (2.2\,nm from \citenum{Hansen2022_radiius}) and assuming half-filled hexagonal packing. We note that considering higher exciton densities participating in strong coupling would lead to a smaller Rabi contraction. We multiply this number by 37, the number of metal-halide quantum wells in a 60-nm film. The number of bleached excitons $N_\text{bl}$ is estimated from the experimental excitation density of 0.7~$\mu$J\,cm$^{-2}$ at a center wavelength 528\,nm, assuming the most drastic scenario in which all pulse energy is absorbed and only causes exciton bleaching. Then, $N_2 = N_1 - N_\text{bl}$. After setting the initial g$_1$ as 90\,meV (from our simulations), the estimated effect of the Rabi contraction would be 2\,meV. The shift is based on an overly estimated density and thus is an upper limit of the possible shift. Notably, even this upper estimate is too small to result in the observed cross-peaks in the rephasing spectrum presented in the manuscript. Accordingly, we can disregard Rabi contraction being the source of the observed nonlinear lineshapes. 

\section*{Gross-Pitaevskii equation with exciton diffusion}
\begin{align}
    i\hbar\partial_t \phi(x) =& \left(-\frac{\hbar^2}{2\mu}\nabla^2 + 
    g|\phi(x)|^2 + \frac{i}{2}(r S(x)-\gamma)\right)\phi(x); \\
    \partial_t S(x) =& \: D_s\nabla^2 S(x) - rS(x)|\phi(x)|^2  
    \nonumber - \gamma_{SS}S^2(x)
    +p(x,t) - \gamma_S S(x),
\end{align}

where $|\phi(x)|^2$ is the condensate population and   
$S(x)$ is the exciton reservoir, pumped by $p(x,t)$, 
which has a characteristic decay rate 
$\gamma_S$ and can interconvert 
with the
condensate population at a bimolecular rate $r$. 
Furthermore, $g$ is the mean-field 
interaction between polaritons and 
$\gamma$ represents the losses of the condensate.  
The model also includes 
exciton-exciton annihilation 
depleting the exciton reservoir population with rate constant $\gamma_{SS}$.  
This model suggests that condensates can form when a steady-state population 
can be established, requiring that $(rS(x)-\gamma) >0$
so that the rate of polariton 
population through the exciton reservoir outpaces
the condensate's population loss.  

\section*{Inelastic Polariton Scattering via Dark Excitonic States}

Inelastic scattering between polariton modes can be described as a second-order perturbative process involving virtual excitation of intermediate dark excitonic states. This is analogous to Raman scattering, 
where an initial polariton state \(|1\rangle\) is scattered into a final state \(|2\rangle\) through a virtual manifold \(|n\rangle\) of dark excitonic modes:
\begin{align}
|1\rangle &\xrightarrow{H_{\text{int}}} \left\{ |n\rangle \right\} \xrightarrow{H_{\text{int}}} |2\rangle, \quad E_n \in \text{continuum above } E_1, E_2.
\end{align}
The transition amplitude for such a process is given by second-order perturbation theory as
\begin{align}
T_{12}^{(2)} = \sum_n \frac{\langle 2 | H_{\text{int}} | n \rangle \langle n | H_{\text{int}} | 1 \rangle}{E_1 + \hbar \omega - E_n + i\eta},
\end{align}
where \(E_1\) is the energy of the initial npolariton state, \(\omega\) is the frequency of the incoming probe field, \(E_n\) are the energies of intermediate dark states, and \(\eta\) encodes the lifetime broadening.

When the intermediate states form a dense continuum, the sum can be replaced by an integral over the dark-state spectral density \(\rho(\omega')\), frequency-dependent light–matter coupling \(g(\omega')\), and a steady-state dark-state population \(n_{\text{SS}}(\omega')\):

\begin{align}
T_{12}(\Delta\omega) = \int d\omega'\, \rho(\omega') \frac{g(\omega')^2 \cdot n_{\text{SS}}(\omega')}{\omega_1 + \Delta\omega - \omega' + i\eta},
\end{align}
where \(\omega_1\) is the energy of the initial polariton state. The population factor \(n_{\text{SS}}(\omega')\) accounts for the occupation of the dark states. In thermal equilibrium, \(n_{\text{SS}}(\omega') = \beta e^{-\beta \omega'}\), which would exponentially suppress transitions involving high-energy intermediate states and effectively kill \(T_{12}\). However, in a non-equilibrium steady state driven by cavity pumping or incoherent excitation, \(n_{\text{SS}}(\omega')\) can be non-thermal. In the absence of a microscopic model, we treat \(n_{\text{SS}}(\omega')\) as a parameterized function.

We define the integrand as a product of a numerator \(f(\omega')\) and a complex-valued denominator \(D(\omega')\), so that:
\begin{align}
T_{12}(\Delta\omega) = \int d\omega'\, \frac{f(\omega')}{D(\omega')} \quad \text{with} \quad D(\omega') = \omega_1 + \Delta\omega - \omega' + i\eta,
\end{align}
and \(f(\omega') = \rho(\omega') g(\omega')^2 n_{\text{SS}}(\omega') \equiv e^{-S(\omega')}\).

To evaluate the integral analytically, we invoke the method of steepest descent. Assume both the density of states and coupling profile are Gaussian and normalized:
\begin{align}
\rho(\omega') = \frac{1}{\sqrt{2\pi}\sigma} \exp\left[-\frac{(\omega' - \omega_e)^2}{2\sigma^2}\right], \quad
g(\omega')^2 = \frac{1}{\sqrt{\pi}\delta} \exp\left[-\frac{(\omega' - \omega_e)^2}{\delta^2}\right].
\end{align}

Combining these with a constant steady-state population, the exponent becomes:
\begin{align}
S(\omega') = \frac{(\omega' - \omega_e)^2}{2\sigma^2} + \frac{(\omega' - \omega_e)^2}{\delta^2}.
\end{align}

The saddle point \(\omega_s\) is determined from \(S'(\omega') = 0\), yielding:
\begin{align}
\omega_s = \omega_e, \quad \text{where} \quad \gamma = \frac{1}{2\sigma^2} + \frac{1}{\delta^2}.
\end{align}
The second derivative \(S''(\omega_s) = 2\gamma\) confirms a local minimum, ensuring the validity of the steepest descent approximation. Physically, \(\omega_s\) corresponds to the most probable intermediate dark-state energy contributing to the scattering process, balancing thermal accessibility and spectral weight. It represents the effective center of entropic support for the transition. The resulting expression for the scattering amplitude is:
\begin{align}
T_{12}(\Delta\omega) \approx \frac{f(\omega_s)}{\omega_1 + \Delta\omega - \omega_s + i\eta} \cdot \sqrt{\frac{2\pi}{S''(\omega_s)}},
\end{align}
where \(f(\omega_s)\) captures the prefactor evaluated at the saddle point, including normalization.
This approximation reveals that the scattering amplitude is dominated by intermediate frequencies near \(\omega_s\), and peaks when \(\omega_1 + \Delta\omega \approx \omega_s\). The entropic width and shift in \(\omega_s\) encode the competition between energy detuning and the structure of the steady-state distribution.

This expression encodes the competition between energy and entropy in the scattering process. While the Boltzmann factor \(e^{-\beta \Delta\omega}\) would thermodynamically favor energy-loss transitions (i.e., \(-\Delta\omega\)) in equilibrium, a steady-state distribution weighted toward higher energy dark states can bias the amplitude toward energy-gain (up-conversion) processes. The result is a balance: when the spectral weight carried by the dark exciton distribution overcomes the energetic suppression of up-scattering, \(T_{12}(+\Delta\omega)\) dominates. This mechanism closely parallels thermally assisted anti-Stokes Raman processes in molecular systems.

\clearpage

\begin{figure}
    \centering
    \includegraphics[width=0.7\textwidth]{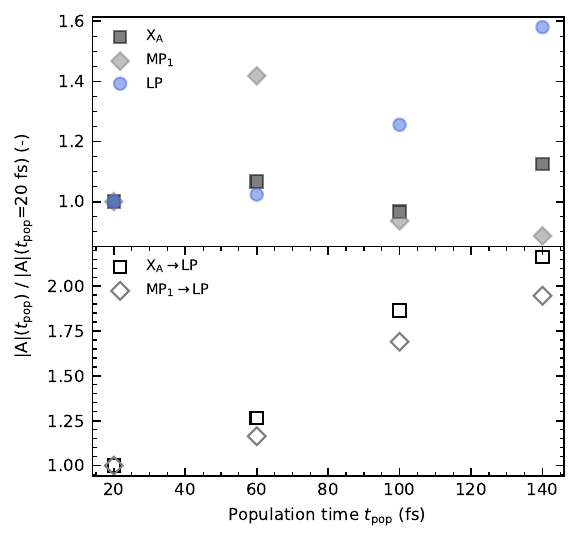}
    \caption{Evolution of the LP, X$_\mathrm{A}$, and MP$_1$ populations (top) and X$_\mathrm{A}\rightarrow$ LP and MP$_1\rightarrow$ LP cross-peaks (bottom) as a function of population time ($t_\mathrm{pop}$). For the populations, the data points correspond to the amplitude ($|\mathrm{A}|$) of the diagonal features at a given $t_\mathrm{pop}$ and, for the cross-peaks, the data points are given by the amplitude of the features in a cut along the emission energy axis at 2.342 eV. The complete diagonal lineshape and cut along the emission energy axis are shown in the main text in Fig.3(h,i).}
    \label{fig:1QAmplitudes_SI}
\end{figure}

\begin{figure}
    \centering
    \includegraphics[width=0.6\textwidth]{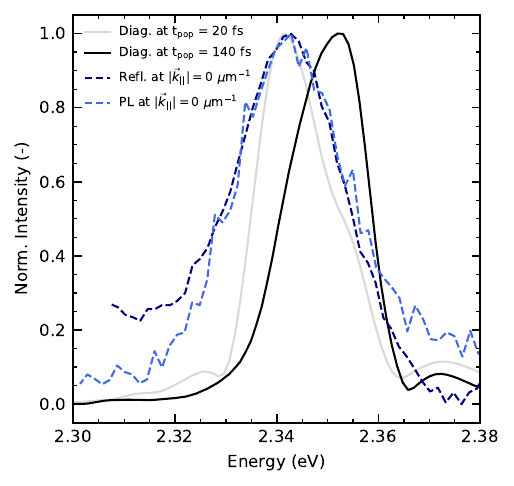}
    \caption{Comparison between the lower polariton features in reflectance and PL at $\vec{k_\parallel}=\vec{0}$ and the diagonal of 1Q spectra at $t_\mathrm{pop}=$ 20, 140 fs.}
    \label{fig:linewidths_SI}
\end{figure}

\begin{figure}
    \centering
    \includegraphics[width=0.7\textwidth]{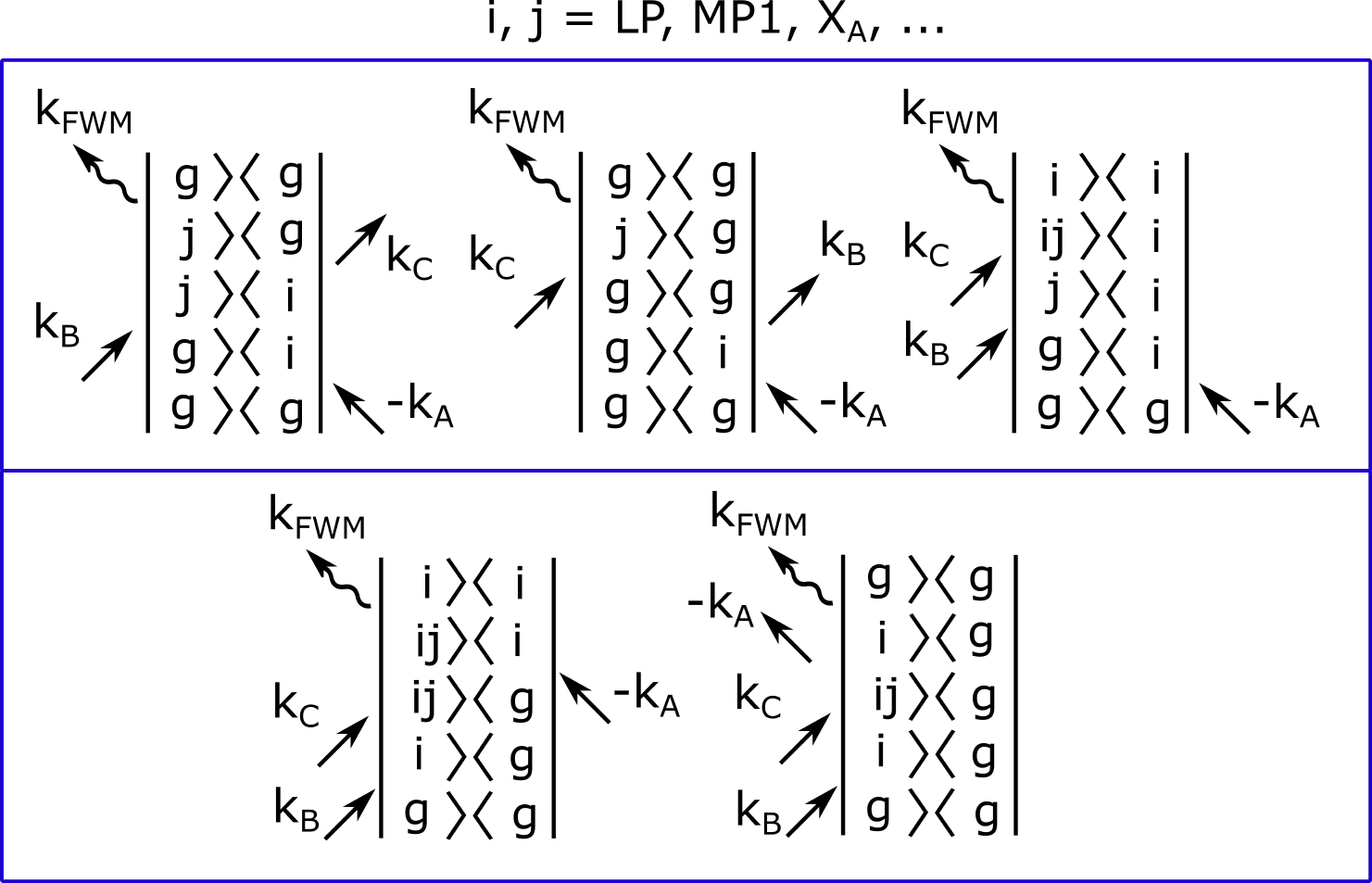}
    \caption{Double-sided Feynman diagrams showing the excitation pathways contributing to the 1Q rephasing (top) and 2Q non-rephasing spectra (bottom).}
    \label{fig:FeynmanSI}
\end{figure}

\begin{figure}[h]
\centering
\includegraphics[width=0.6\textwidth]{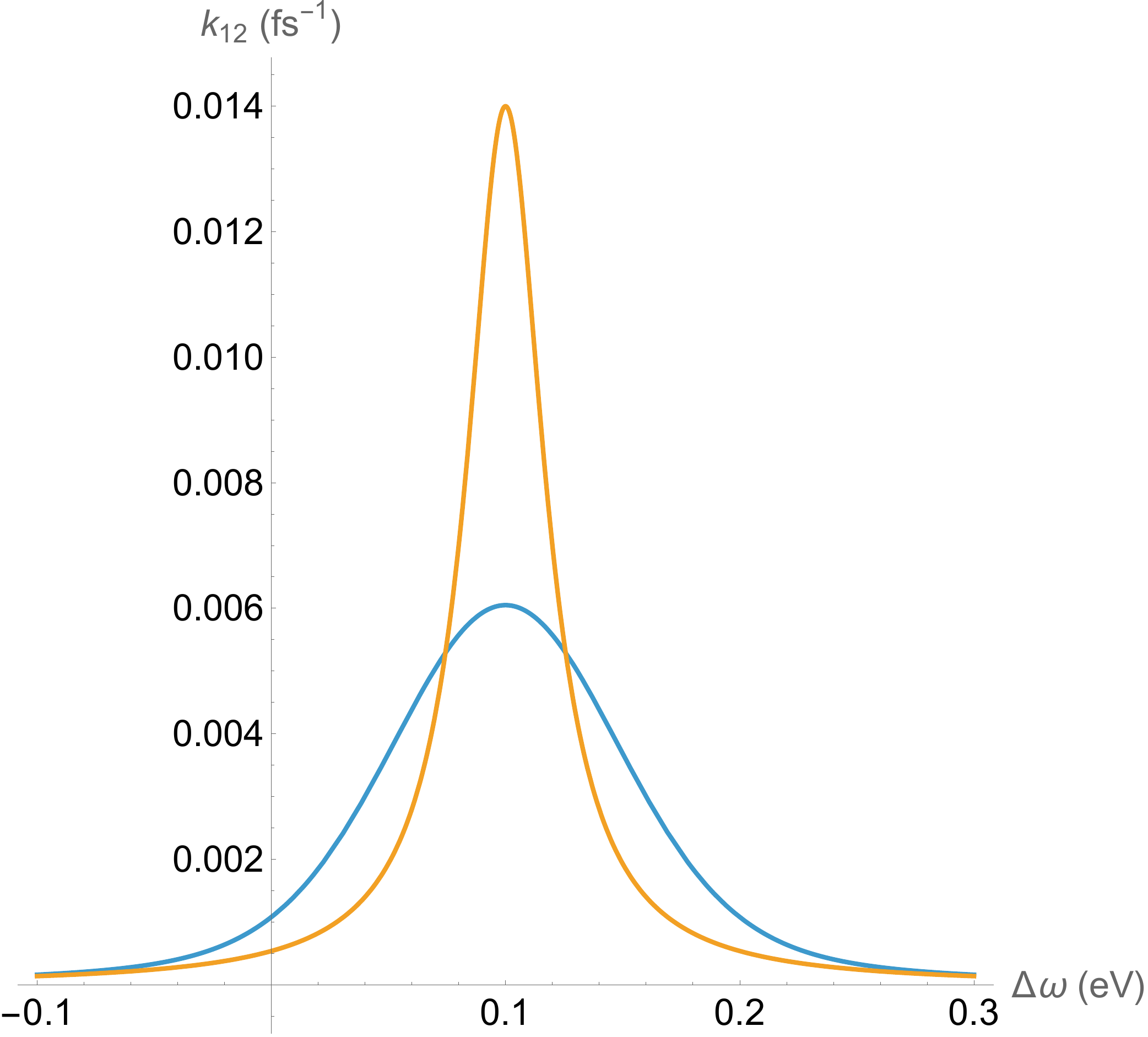}
\caption{
Comparison of exact (blue) and steepest descent (orange) evaluations of the inelastic scattering rate  as a function of energy detuning. The sharper orange curve illustrates the approximation focused around the dominant saddle point contribution, while the broader blue curve results from direct numerical integration. Agreement near the resonance confirms the validity of the steepest descent method in the low-temperature, weakly broadened regime.  Parameters: average dark-state energy: $\hbar\omega_e = 2.1$\,eV;  peak coupling: $g_{o} = 0.02$\,eV;  Dark state DOS width $\sigma = 0.05$\,eV; coupling width: $\delta = 0.1$\,eV, and initial polariton frequency: $\hbar\omega_{1} = 2.0$\,eV.
}
\label{fig:scatteringRates}
\end{figure}







\end{document}